\documentclass{kluwer}    % Specifies the document style.
\usepackage{graphicx}

\newdisplay{guess}{Conjecture}

\begin{document}
\begin{article}
\begin{opening}
\title{Updating Seismic Hazard at Parkfield}
\author{\'Alvaro \surname{Gonz\'alez}\email{alvaro.gonzalez@unizar.es}\thanks{Author for correspondence.}}
\institute{Departamento de Ciencias de la Tierra, \\ Universidad
de Zaragoza, C.~Pedro Cerbuna,~12, 50009 Zaragoza, Spain}
\author{Javier B. \surname{G\'omez}\email{jgomez@unizar.es}}
\institute{Departamento de Ciencias de la Tierra, \\ Universidad
de Zaragoza, C.~Pedro Cerbuna,~12, 50009 Zaragoza, Spain}
\author{Amalio F.
\surname{Pacheco}\email{amalio@unizar.es}} \institute{Departamento
de F{\'i}sica Te\'orica  and BIFI, \\ Universidad de Zaragoza,~C.
Pedro Cerbuna,~12, 50009 Zaragoza, Spain}
\runningauthor{Gonz\'alez et al.} \runningtitle{Updating Seismic
Hazard at Parkfield}

\begin{abstract}
The occurrence of the September 28, 2004 $M_w$=6.0 mainshock at
Parkfield, California, has significantly increased the mean and
aperiodicity of the series of time intervals between mainshocks in
this segment of the San Andreas fault. We use five different
statistical distributions as renewal models to fit this new series
and to estimate the time-dependent probability of the next
Parkfield mainshock. Three of these distributions (lognormal,
gamma and Weibull) are frequently used in reliability and
time-to-failure problems. The other two come from physically-based
models of earthquake recurrence (the Brownian Passage Time Model
and the Minimalist Model). The differences resulting from these
five renewal models are emphasized.

\end{abstract}
\keywords{California, Parkfield, renewal models, San Andreas
fault, seismic hazard assessment\\\\
\bf{Accepted for publication in Journal of Seismology.}}

\end{opening}

\section{Introduction}
Renewal models are frequently used to estimate the long-term
time-dependent probability of the next large earthquake on
specific faults or fault segments where large shocks occur
repeatedly at approximately regular intervals. In this approach,
it is assumed that the times between consecutive large earthquakes
(inter-event times or recurrence intervals) follow a certain
statistical distribution. The available record of large
earthquakes in any given fault is typically scarce (usually
including less than ten events), so the empirical statistics are
poor. Because of this, a theoretical distribution is fitted to the
observed inter-event times and used to estimate future earthquake
probabilities. In the first instance, several well-known
statistical distributions were used \cite{Utsu,Ellsworth}. These
distributions have only an empirical rooting and are commonly
employed as renewal models in reliability and time-to-failure
problems \cite{Ellsworth}. Three of these distributions (the
gamma, lognormal, and Weibull) are frequently used for
earthquakes, because they share three properties commonly observed
for earthquake inter-event times \cite{Michael}: First, these
times must be positive, and these distributions only exist for
positive times; second, inter-event times much smaller than the
average recurrence interval are rare; and third, the distribution
of inter-event times decays slowly for times longer than the
average.

More recently, the distributions derived from two simple physical
models of earthquake recurrence have been proposed as an
alternative to those purely empirical approaches. These two models
have the virtue of providing an intuitive picture of the seismic
cycle in a fault or fault segment. They are the Brownian Passage
Time Model \cite{KaganKnopoff,Ellsworth,Matthews,WGCEP}, and the
Minimalist Model \cite{VP,GomezAndPacheco}. The first one
represents the tectonic loading of a fault by a variable which
evolves by superposition of an increasing linear trend and a
Brownian noise term, and an earthquake occurs when this variable
reaches a given threshold \cite{Matthews}. All the earthquakes in
this model are identical to each other. The Minimalist Model
sketches the plane of a seismic fault, where earthquake ruptures
start and propagate according to simplified breaking rules. This
model generates earthquakes of various sizes, and only the time
between the largest ones (the characteristic earthquakes, that
break the whole model fault) are considered for the inter-event
time distribution. The distributions derived from these two
models, as well as the gamma, lognormal and Weibull, generally
represent fairly well the observed distribution of
large-earthquake inter-event times \cite{GomezAndPacheco}.
However, they differ significantly in their probability
predictions for times much longer than the mean inter-event time
of the data. Thus, it seems convenient to take all their different
predictions into account.

The paucity of data in the record of large earthquakes in a given
fault or fault segment has another consequence. The occurrence of
a new large shock significantly modifies the mean and aperiodicity
of the available series of inter-event times. In turn, this
significantly modifies the parameters of the best fit of the
statistical distribution used for the adjustment, resulting in
different estimations of future earthquake probabilities. Thus,
for seismic hazard assessment, it is important to recalculate the
new parameters. This is just the case at the Parkfield segment of
the San Andreas fault in California with the September 28, 2004,
$M_w=6.0$ earthquake, which significantly increased both the mean
and the aperiodicity of the available series. Therefore, the
purpose of this short communication is to update the previous fits
\cite{Ellsworth,GomezAndPacheco} to this new series and compare
the hazard predictions coming from five different renewal models:
the gamma (G), lognormal (LN) and Weibull (W) distributions as
classical renewal models, and also the Brownian Passage Time Model
(BPT) and the Minimalist Model (MM), as more recent counterparts.

In Section II, the mean and the aperiodicity of the new series are
calculated.  Besides, this section contains the best fits obtained
by the method of moments with the different renewal models.
Finally, in Section III we discuss the probability estimates for
the next mainshock at Parkfield.

\section{Fits to the new series}
Including the latest event, the Parkfield series
\cite{Bakun,Bakun2,MichaelJones} consists of seven $M_w\simeq6$
mainshocks, which occurred on January 9, 1857; February 2, 1881;
March 3, 1901; March 10, 1922; June 8, 1934; June 28, 1966 and
September 28, 2004. In consequence, the duration (in years) of the
six observed inter-event times are: 24.07, 20.08, 21.02, 12.25,
32.05 and 38.25. The mean value $m$, the sample standard deviation
$s$ (the square root of the bias-corrected sample variance), and
the aperiodicity $\alpha$ (equivalent to the coefficient of
variation, i.e. the standard deviation divided by the mean) of
this six-data series are:
\begin{equation}\label{values}
m = 24.62 \mbox{ yr} \qquad s = 9.25 \mbox{ yr} \qquad \alpha =
0.3759
\end{equation}

Now, we will proceed to fit these data using the G, LN and W
families of distributions \cite{Utsu} and the BPT and MM models
\cite{Matthews,GomezAndPacheco}. The statistical distribution of
inter-event times in the BPT model is the so-called inverse
Gaussian distribution, which, as the three classical distributions
mentioned at the beginning, is a continuum biparametric density
distribution. Strictly speaking, the distribution derived from the
MM is a discrete one and has only one parameter, $N$ (the number
of cells in which the model fault plane is divided, directly
related to the aperiodicity $\alpha$ of the series)
\cite{GomezAndPacheco}. However, for fitting the data, it is
necessary to assign a definite number of years to the
non-dimensional time step of the model. This second parameter will
be called $\tau$. The distribution in the MM has an analytical
solution, explicitly written for $N\le5$ in
\citeauthor{GomezAndPacheco} (\citeyear{GomezAndPacheco}). For
larger values of $N$ the distribution can be calculated
numerically with Monte Carlo simulations.

Next, we will write down the explicit analytic form of the four
mentioned continuum probability density distributions. Each of
them has a scale parameter and a shape parameter.
In all formulae the time, $t$, is measured in years.\\
\\
\noindent Gamma distribution:

\begin{equation}
G(t)=\frac{c}{\Gamma(r)}(ct)^{r-1}e^{-ct},\quad c>0,\; r>0
\end{equation}

\noindent Lognormal distribution:

\begin{equation}
LN(t)=\frac{1}{\sqrt{2\pi}\sigma t}\exp\left[-\frac{(\ln t -
n)^2}{2\sigma^2}\right],\qquad n>0,\; \sigma>0
\end{equation}

\noindent Weibull distribution:

\begin{equation}
W(t)=a\rho t^{\rho -1}\exp(-at^\rho),\qquad a>0,\; \rho>0
\end{equation}

\noindent Brownian Passage Time distribution \cite{Matthews}:

\begin{equation}
BPT(t)=\left(\frac{m}{2 \pi \alpha^2
t^3}\right)^{1/2}\exp\left[-\frac{(t-m)^2}{2m\alpha^2t}\right]
\end{equation}

\noindent In this last case, the parameters $m$ and $\alpha$
correspond to the mean and aperiodicity defined earlier.

We will use the method of moments to fit the data, so within these
four families of distributions, and the same for the MM, we will
select that specific distribution whose mean value and
aperiodicity are equal to those of the Parkfield series
(Eq.~\ref{values}). The specific values of the parameters that
make the different distributions fulfill this condition are
written in Table \ref{parset}.

\begin{table}[h] %
\begin{tabular}{lc}
\hline
Gamma      & $c=0.287 \; yr^{-1}\quad  r=7.078 $\\
Lognormal     & $n=3.137 \quad \sigma=0.364$\\
Weibull     & $a=6.853\times 10^{-5} \; yr^{-\rho} \quad \rho=2.889$\\
BPT   & $m=24.62 \; yr \quad \alpha=0.3759 $\\
MM   & $N=495 \quad (\alpha=0.3759) \quad \tau=4.186 \times
10^{-3}\; yr$\\
\hline
\end{tabular}
\caption[]{Parameter values obtained by the method of moments for
the five renewal models described in the text for the Parkfield
series.}\label{parset}
\end{table}

Note that in the MM model the aperiodicity of the series fixes the
value of the parameter $N$ \cite{GomezAndPacheco}. In a minimalist
system with $N=495$, the mean recurrence interval of the
characteristic earthquakes is 5881.2 non-dimensional time steps.
Comparing this mean with the value $m = 24.62$ yr quoted in
Eq.~\ref{values}, we deduce that one time step of the model
corresponds to $\tau = 24.62$ yr$/5881.2 = 4.186 \times 10^{-3}$
yr, or around 1.5 days.

In Fig. 1a, we have superimposed the cumulative histogram
(empirical distribution function) of the Parkfield series together
with the cumulative distributions of the five models. These, for
the G, LN, W and BPT are obtained by integrating Eqs. 2-5, and for
the MM by summing its discrete probability distribution. In Fig.
1b we show the residuals for the five fits. Finally, in Fig. 2 we
present the annual (conditional) probability of occurrence derived
from the five models \cite{Utsu}.

\begin{figure}[h]
\includegraphics[width=12cm]{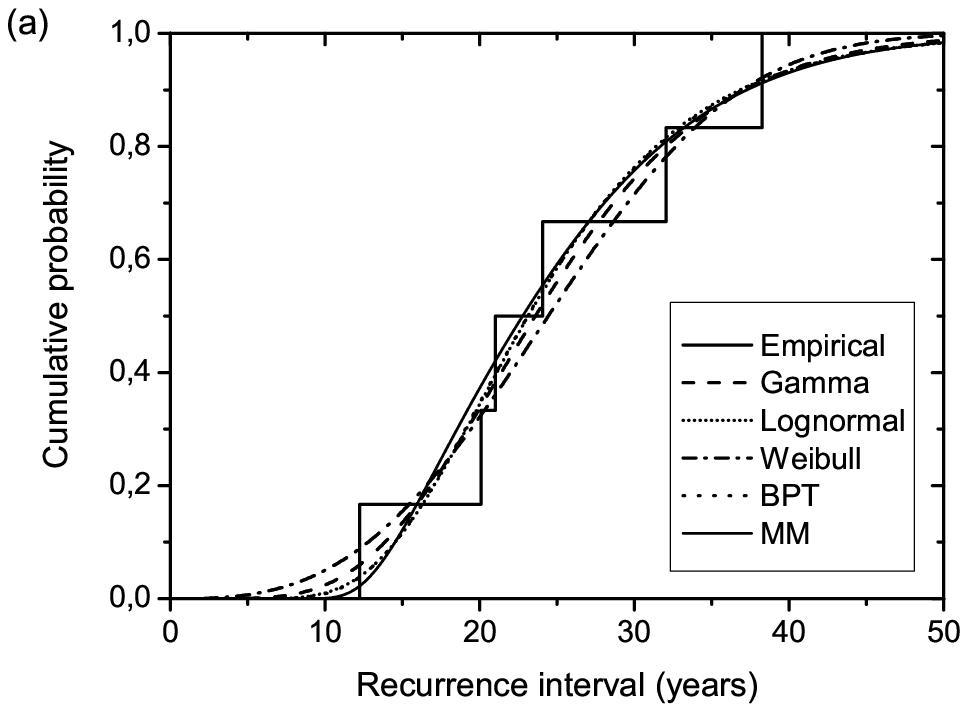}
\includegraphics[width=12cm]{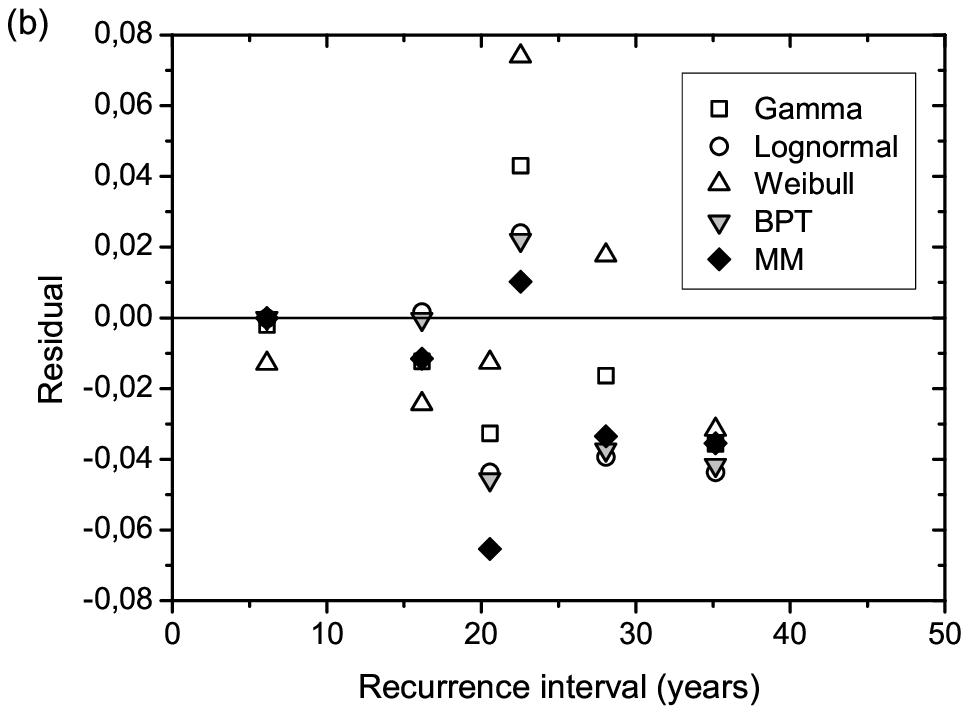}
\caption[]{\label{fig:1}Comparison of the performance of different
renewal models. (a) Fit to the Parkfield sequence of the gamma,
lognormal, Weibull, BPT and MM models; (b) residuals for the five
model fits, evaluated at the midpoints of the horizontal segments
of the empirical distribution function.}
\end{figure}

\begin{figure}[h]
\includegraphics[width=12cm]{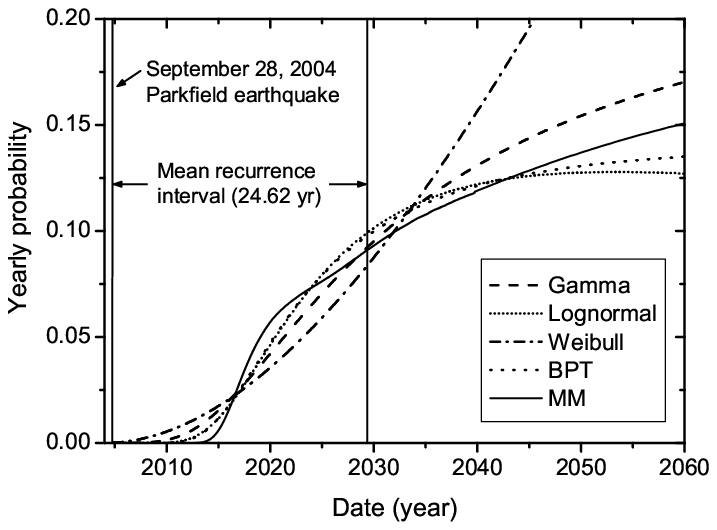}
\caption[]{\label{fig:2}Annual probability for the new mainshock
at Parkfield according to the five renewal models.}
\end{figure}

\section{Discussion and conclusions}

The results shown in Fig. 1a and 1b indicate that the five models
used in the adjustment describe rather well the Parkfield data,
despite being different from each other. All the residuals (Fig.
1b) are lower than 8\% at any time (most of them lower than 4\%).
The minimum residual is not always related to the same model, and
the `best' model changes with time. This precludes selecting one
particular model as the optimal choice for prediction purposes.

It is clear from Fig. 1a that the curve corresponding to the MM
takes off later than the others. This is because in this model
there exists an initial stress shadow with a length of $N=495$
time steps, i.e. $495\times4.186\times10^{-3}$ yr, that is 2.07
yr. Thus, in this model, in the initial 2.07 yr of the cycle
definitely no new event can occur. This is in stark contrast to
the other renewal models, where there is no strict stress shadow.
Most distributions have a low or very low probability of
occurrence between 0 an 2.07 yr, but none have a strictly zero
probability as the MM has. The LN and BPT curves plot one upon the
other in Fig. 1a because in the time range shown in the graph
their cumulative probability distributions are similar. Note also
that the Weibull model predicts a cumulative probability
considerably higher for $t<10$ yr than the other four models, and
that all five cumulative distribution functions appear to
`converge' in probability roughly 18 yr after the last mainshock,
with a cumulative probability of around 25\%. All the models
indicate that an immediate rerupture of the Parkfield segment is
unlikely (the cumulative probabilities are lesser than 1\% for the
first five years), but it will likely occur not later than 53
years after the last one (moment at which all cumulative
probabilities are at least 99\%).

In Fig. 2, where the annual probability of occurrence is shown,
there are several observations worth comment. At the beginning of
the cycle the W curve is the first in the take off and the MM is
the last. This reflects what was mentioned in the previous
paragraph. Later, there is an interval, roughly speaking from 2016
to 2023, in which the MM curve is on top of the others, predicting
slightly higher annual probabilities.  Around 2030, when the mean
recurrence interval of the series has elapsed, all the models
predict a yearly conditional probability between 8.5\% and 10\%.
The predictions from the LN and the BPT are very similar until
about 2040, because the BPT distribution is similar to, but not
identical to, a lognormal distribution \cite{Michael}. But from
2040 onwards, the five models start showing their asymptotic
behaviour, or, in other words, their clear discrepancies. The
behaviour of the conditional probability for long times after a
large earthquake is still debated (e.g. \citeauthor{Davis} 1989;
\citeauthor{Sornette}, 1997), and the different models used in
this paper show three different possibilities for it. First, a
decreasing probability, the case of the LN model, where the
probability starts declining from the year 2053 onwards,
approaching zero as time passes. Second, a probability which
increases asymptotically to 100\%, the case of the W model,
according to which there is a 95\% yearly probability of having a
mainshock after 163 yr from the last one. And third, a probability
that increases towards an asymptote smaller than 100\%, the case
of the curves predicted by the BPT, G and MM models. The
asymptotic yearly probability value is different for each of these
three models: 13\%, 26\%, and 38\% respectively. This last value
in the MM is given by the formula
\begin{equation}
\lim_{n\to\infty}  P(n,\tau) =
\frac{N-1}{N}\left[1-\left(\frac{N-1}{N}\right)^\frac{1}{\tau}\right],
\end{equation}
where $n$ is the number of time steps since the last characteristic
earthquake in the model.

The discrepancies between the predictions of these five approaches
cannot be used to disregard any of them, at least for the first
decades since the last mainshock. On the contrary, they can be
considered all together to give reasonable upper and lower bounds
to the annual probability of occurrence at Parkfield: between
8.5\% and 10\% after 25 yr (i.e., after one mean cycle length),
and between 12\% and 17\% after 37 yr (i.e., after 1.5 mean cycle
lengths).

\acknowledgements This research is funded by the Spanish Ministry
of Education and Science, through the project BFM2002--01798, and
the research grant AP2002--1347 held by \'AG.

\end{article}
\end{document}